\documentstyle[multicol,aps]{revtex}
\begin{document}
\title{Conformal symmetry and quantum relativity}
\author{Marc-Thierry Jaekel$^a$ and Serge Reynaud$^b$}
\address{$(a)$ Laboratoire de Physique Th\'{e}orique de l'Ecole
Normale Sup\'{e}rieure,\\
Centre National de la Recherche Scientifique, Universit\'{e} Paris Sud, \\
24 rue Lhomond, F75231 Paris Cedex 05 France\\
$(b)$ Laboratoire Kastler Brossel, Universit\'{e} Pierre et Marie Curie,
case 74\\
Ecole Normale Sup\'{e}rieure, Centre National de la Recherche Scientifique,
\\
4 place Jussieu, F75252 Paris Cedex 05 France}
\date{LPTENS 97/23}
\maketitle

\begin{abstract}
The relativistic conception of space and time is challenged by
the quantum nature of physical observables.
It has been known for a long time that Poincar\'e symmetry
of field theory can be extended to the larger conformal symmetry.
We use these symmetries to define quantum observables associated with
positions in space-time, in the spirit of Einstein theory of
relativity.
This conception of localisation may be applied to
massive as well as massless fields.
Localisation observables are defined as to obey Lorentz
covariant commutations relations and in particular include
a time observable conjugated to energy.
Whilst position components do not commute
in presence of a non-vanishing spin,
they still satisfy quantum relations which generalise
the differential laws of classical relativity.
We also give of these observables
a representation in terms of
canonical spatial positions, canonical spin components
and a proper time operator conjugated to mass.
These results plead for a new representation not only of
space-time localisation but also of motion.
\end{abstract}

\begin{multicols}{2}

\section{Introduction}

Space and time are now considered as closely linked to each other. This is a
consequence of relativistic conceptions of space-time which rely on the
notion of events localised both in space and time \cite{Einstein05}. Time
comparison between remote clocks is performed through synchronisation
procedures built on the transfer of electromagnetic signals and distance
measurement is performed through localisation procedures built on two-way
transfer. These ideas are now included in metrological definitions of time
and space units as well as in a number of practical applications \cite
{TimeFrequency}. It is worth emphasizing that the space and time associated
with an event are physical observables delivered by specifically designed
apparatus. In particular, the time associated with a given event does not
evolve and must therefore be clearly distinguished from any evolution
parameter which may be used to write dynamical laws and conservation laws.
At the same time, the physical observables describing space-time positions
cannot be confused with classical coordinate parameters on a space-time map.
Their definition has to reach limits associated with the quantum nature of
the physical world \cite{SaleckerWigner}. More profoundly, these observables
certainly belong to the quantum domain, like atomic clocks used for time
definition and electromagnetic signals used for synchronisation or
localisation.

The problem of space-time localisation however raises challenging issues in
the context of standard quantum formulations. The definition of space
positions is delicate in the presence of spin \cite
{Schrodinger,Pryce48,NewtonWigner,Fleming65,BarutMalin}. Moreover, time is
usually treated as a classical parameter rather than as an operator \cite
{Jammer74} and the difference in the description of space and time variables
leads to considerable difficulties in the attempts to build quantum and
relativistic theories of gravity \cite{Rovelli91,Isham93}. More generally,
the representation of evolution remains a challenge in any quantum framework
where the prime roles are played by conserved quantities \cite
{PageWootters,Unruh89,Rovelli95}. In the present paper, we show that
attaching more importance to the symmetry properties of the physical theory
allows to progress towards a solution of these difficulties.

The advent of relativity theory was mainly based upon the symmetry of
electromagnetism under Lorentz transformations. Connecting localisation in
space-time to propagation of electromagnetic pulses, Einstein was able to
derive the relativistic transformations of space and time variables.
He then discovered the law of inertia of energy as a
consequence of conservation of the generators associated with Lorentz boosts
\cite{Einstein06}. Shortly after the birth of relativity, Bateman and
Cunningham showed that Lorentz symmetry of electromagnetism can be extended
to the larger group of conformal coordinate transformations \cite{BC09}
which preserve Maxwell equations while fitting the relativistic definition
of uniformly accelerated motion. The conservation of generators associated
with this symmetry was soon established \cite{BesselHagen}. Conformal
symmetry implies that the propagation of electromagnetic fields is not
sensitive to a conformal variation of the metric tensor, that is a change of
space-time scales preserving the velocity of light \cite{MG80}.

The interpretation of conformal symmetry as a general symmetry of the laws
of physics has exerted an obvious attraction on a number of physicists but
it received at the same time severe objections from other ones and sometimes
from the same ones \cite{Pauli21,Page36,Dirac36,Hill45,Gupta61,FRW62,BFH83}.
A recurrent matter of debate is the physical significance of the conformal
generators. Another one is the pertinence of conformal symmetry for massive
fields or massive objects. It has been known for long that mass has
to vary under conformal transformations so that massive field theories have
be modified to remain invariant \cite{SchoutenHaantjes,BarutHaugen}. The
modification reduces to a transformation of the mass parameter appearing in
the equations, which simply means that mass scales vary as reciprocals of
space-time scales under conformal transformations. Clearly, this behaviour
has to be expected from the dimensional relation which connects mass and
space-time as soon as the velocity of light $c$ and the quantum constant $%
\hbar $ are treated as fundamental constants \cite
{BarutHaugen,Dicke,Sakharov,Hoyle}. Despite these attractive features and
the lasting efforts of many physicists, among which those of Barut \cite
{BarutConformal}, a full recognition of the value of conformal symmetry as
an enlarged form of Lorentz symmetry has not already been reached.

The purpose of the present paper is to convince the reader
that conformal symmetry is an appropriate
tool for addressing the questions raised in the beginning of the
Introduction. Conformal symmetry allows to give definitions of space-time
observables associated with an event and to show that these definitions
fulfill satisfactory quantum and relativistic properties \cite
{PRL96,PLA96,EPL97}. Space-time observables are built on
the conformal generators, that is the conserved quantities associated with
conformal symmetry. The conformal algebra, that is the commutators of
conformal generators, not only fixes the quantum commutators of observables
but also their relativistic shifts under changes of reference frames. In
particular, the shift of mass under transformations to accelerated frames
may be read as defining the space-time observables, in consistency with the
redshift of energy and momentum known from classical relativity \cite
{Einstein07}.

When a specific quantum event like the annihilation of an electron-positron
pair into a pair of photons is considered, it becomes clear that the
space-time position of the annihilation event may be defined, roughly
speaking, as the point of coincidence of the two emitted electromagnetic
pulses. This definition thus enters the general framework of space-time
localisation which has been outlined previously. It nevertheless applies to
massive objects such as electron and positron present before the
annihilation as well as to massless objects such as the emitted photons, as
a direct consequence of conservation laws. The formalism presented in the
present paper will allow us to account quantitatively for the various
dispersions associated with the quantum nature of the event.

We will then propose further arguments pleading for the physical
significance of conformal symmetry. We will use transformation properties of
space-time observables to define a conformal factor which plays the role of
a metric factor while pertaining to the world of quantum observables. We
will also exhibit an equivalent representation of space-time observables in
terms of canonical variables.

\section{Conformal algebra}

To introduce the algebra of conformal transformations, it is convenient to
consider general deformations of a coordinate map which represent
changes of frame as Lie transformations
\begin{equation}
x^\mu \rightarrow x^\mu +\delta ^\mu (x)  \label{map}
\end{equation}
where $\delta ^\mu $ are polynomial functions of coordinate parameters $x$.
A commutator between two deformations $a$ and $b$ may be introduced as the
difference between the composed deformations $a\circ b$ and $b\circ a$. The
difference between the images of a point $x$ through the two deformations is
determined by the Lie commutator
\begin{equation}
\delta _{(a,b)}^\mu =\delta _b^\nu \frac{\partial \delta _a^\mu }{\partial
x^\nu }-\delta _a^\nu \frac{\partial \delta _b^\mu }{\partial x^\nu }
\label{comm}
\end{equation}
Transformations are performed around an inertial frame, so that tensor
indices are raised or lowered by using the Minkowski tensor
\begin{equation}
\eta _{\mu \nu }={\rm diag}\left( 1,-1,-1,-1\right)
\end{equation}
We consider here the transformations corresponding respectively
to translations $P_\nu $, rotations $J_{\nu \rho }$, dilatation $D$ and
conformal transformations to uniformly accelerated frames $C_\nu $
\begin{eqnarray}
\delta _{P_\nu }^\mu (x) &=&\eta _\nu ^\mu  \nonumber \\
\delta _{J_{\nu \rho }}^\mu (x) &=&\eta _\nu ^\mu x_\rho -\eta _\rho ^\mu
x_\nu  \nonumber \\
\delta _D^\mu (x) &=&x^\mu  \nonumber \\
\delta _{C_\nu }^\mu (x) &=&2x_\nu x^\mu -\eta _\nu ^\mu x_\rho x^\rho
\label{cct}
\end{eqnarray}
$\eta _\nu ^\mu $ denotes a Kronecker symbol.

In quantum field theory, the generators of conformal transformations
identify with integrals over a space-like surface of stress tensor
components weighted by functions (\ref{cct}) representing map deformations
\cite{ItzyksonZuber}.
These generators are quantum observables associated with a given field state
and defined in such a manner that they vanish in vacuum. Such a definition
is allowed by the conformal invariance of vacuum \cite{QSOBJP95}. The
generators are conserved quantities which are used in the following to
characterise field states. Their commutators are consistent with Lie
commutators (\ref{comm}) and they therefore determine the quantum
commutation relations as well as the relativistic shifts of field states
under frame transformations \cite{deWitt63}. The conformal algebra is
described as the set of these commutators
\begin{eqnarray}
&&\left( P_\mu ,P_\nu \right) =0  \nonumber \\
&&\left( J_{\mu \nu },P_\rho \right) =\eta _{\nu \rho }P_\mu -\eta _{\mu
\rho }P_\nu   \nonumber \\
&&\left( J_{\mu \nu },J_{\rho \sigma }\right) =\eta _{\nu \rho }J_{\mu
\sigma }+\eta _{\mu \sigma }J_{\nu \rho }-\eta _{\mu \rho }J_{\nu \sigma
}-\eta _{\nu \sigma }J_{\mu \rho }  \nonumber \\
&&\left( D,P_\mu \right) =P_\mu   \nonumber \\
&&\left( D,J_{\mu \nu }\right) =0  \nonumber \\
&&\left( P_\mu ,C_\nu \right) =-2\eta _{\mu \nu }D-2J_{\mu \nu }  \nonumber
\\
&&\left( J_{\mu \nu },C_\rho \right) =\eta _{\nu \rho }C_\mu -\eta _{\mu
\rho }C_\nu   \nonumber \\
&&\left( D,C_\mu \right) =-C_\mu   \nonumber \\
&&\left( C_\mu ,C_\nu \right) =0  \label{ConfAlg}
\end{eqnarray}
The notation $\left( \Delta _a,\Delta _b\right) $ is taken from Lie
commutators (\ref{comm}). Quantum commutators $\left[ \Delta _a,\Delta
_b\right] $ are given for any generators $\Delta _a$ and $\Delta _b$, and
more generally for any observables, by the correspondance rule
\begin{equation}
\left( \Delta _a,\Delta _b\right) \equiv \frac 1{i\hbar }\left[ \Delta
_a,\Delta _b\right]   \label{Comm}
\end{equation}
Double commutators obey the Jacobi identity
\begin{equation}
\left( \left( \Delta _a,\Delta _b\right) ,\Delta _c\right) =\left( \Delta
_a,\left( \Delta _b,\Delta _c\right) \right) -\left( \Delta _b,\left( \Delta
_a,\Delta _c\right) \right)   \label{Jacobi}
\end{equation}
A dot will denote the symmetrised product of operators
\begin{equation}
\Delta _a\cdot \Delta _b\equiv \frac 12\left\{ \Delta _a\Delta _b+\Delta
_b\Delta _a\right\}
\end{equation}

\section{Space-time observables}

As discussed in the Introduction, localisation of an event in space-time is
performed using a field state which, like the field state produced by
annihilation of an electron-positron pair, contains photons propagating
in at least two different directions. Using the energy-momentum variables $%
P_\mu $ associated with this field state, we define a non vanishing mass $M$
from the Lorentz invariant
\begin{equation}
M=\sqrt{\eta ^{\mu \nu }P_\mu P_\nu }=\sqrt{P_\mu P^\mu }  \label{defM}
\end{equation}
Energy-momentum variables $P_\mu $ are conserved in the annihilation process
and characterize the state of the electron-positron pair existing before the
annihilation as well as that of the two-photon state produced by the
annihilation. As a consequence, the mass is also conserved in the process.
This is also true for the emission-absorption processes considered by
Einstein in his derivation of inertia of energy \cite{Einstein06}.

Space-time observables $X_\mu $, associated with the localisation in
space-time of the event, may be built up on the translation, rotation and
dilatation generators \cite{PLA96}
\begin{eqnarray}
X_\mu =J_{\nu \mu }\cdot \frac{P^\nu }{M^2}+D\cdot \frac{P_\mu }{M^2}
\label{defX}
\end{eqnarray}
In a simple semi-classical approach, the two photons may be thought as
point-like energy distributions propagating along straight lines
representing idealised light rays. These two rays are thus considered to
intersect each other at the space-time position of the annihilation event.
In this context, the observables (\ref{defX}) may be identified with the
point of intersection of the two light pulses \cite{PLA96}. This description
is consistent with a classical conception of localisation of events in
space-time \cite{Einstein05}.

Equation (\ref{defX}) provides a more general definition of localisation in
space-time which is in particular compatible with the dispersions associated
with the quantum nature of the event. Clearly, electron and positron cannot
be considered as point-like structures and they bear spin. The emitted
photons obey the laws of diffraction and cannot be identified with
dimensionless points propagating along idealised light rays. As a
consequence, the position of the annihilation event, defined by the
coincidence of the two photons, is a fuzzy spot with a dimension of the
order of Compton wavelength rather than a sizeless point. The definition (%
\ref{defX}) leads to a fully quantum description of space-time localisation.
In more formal words, such a description involves the enveloping field
associated with conformal symmetry, that is the space of rational
expressions of conformal generators. The observables $X_\mu $ are defined in
the enveloping field while forthcoming results are obtained through direct
computation in this space.

Using the Jacobi identity (\ref{Jacobi}), we show from conformal algebra (%
\ref{ConfAlg}) that the observables $X_\mu $ are shifted under translations,
dilatation and rotations exactly as expected from classical relativity
\begin{eqnarray}
&&\left( P_\nu ,X_\mu \right) =-\eta _{\mu \nu }  \nonumber \\
&&\left( D,X_\mu \right) =-X_\mu   \nonumber \\
&&\left( J_{\mu \nu },X_\rho \right) =\eta _{\nu \rho }X_\mu -\eta _{\mu
\rho }X_\nu   \label{PX}
\end{eqnarray}
The first result means that components of positions obey canonical
commutation relations with momenta. In contrast to the situation encountered
in standard interpretations of the quantum formalism \cite{Jammer74}, a time
operator has now been defined and energy-time besides momentum-space
canonical commutation relations have been obtained. Furthermore, these
relations satisfy an explicit Lorentz covariance. This result, as well as
the two other ones, convincingly pleads for the interpretation of observables
$X_\mu $ as positions in space-time associated with a quantum event.

Using the position observables (\ref{defX}), it is possible to write angular
momentum components $J_{\mu \nu }$ as sums of external contributions which
have their usual form in terms of momenta and positions and of internal
observables $S_{\mu \nu }$
\begin{eqnarray}
&&J_{\mu \nu }=P_\mu \cdot X_\nu -P_\nu \cdot X_\mu +S_{\mu \nu }  \nonumber
\\
&&S_{\mu \nu }=\epsilon _{\mu \nu \rho \sigma }S^\rho \frac{P^\sigma }M
\nonumber \\
&&S^\mu =-\frac 12\epsilon ^{\mu \nu \rho \sigma }J_{\nu \rho }\frac{%
P_\sigma }M  \label{defS}
\end{eqnarray}
The vector $S_\mu $ is the Pauli-Lubanski vector, a covariant generalisation
of spin, while $\epsilon _{\mu \nu \lambda \rho} $ is the antisymmetric
Lorentz tensor \cite{ItzyksonZuber}.
Spin is transverse with respect to energy-momentum
\begin{equation}
P^\mu S_{\mu \nu }=P_\mu S^\mu =0  \label{trans}
\end{equation}
and is invariant under translations and dilatation while being rotated as a
vector under rotations
\begin{eqnarray}
&&\left( P_\mu ,S_\rho \right) =\left( D,S_\rho \right) =0  \nonumber \\
&&\left( J_{\mu \nu },S_\rho \right) =\eta _{\nu \rho }S_\mu -\eta _{\mu
\rho }S_\nu
\end{eqnarray}
Spin components obey simple commutation relations
\begin{equation}
\left( S_\mu ,S_\nu \right) =S_{\mu \nu }
\end{equation}

The redshift laws for energy-momentum no longer have a classical form in
presence of non vanishing spin. Whereas $D$ has a classical form in terms of
momenta and positions (\ref{defX})
\begin{equation}
D=P^\rho \cdot X_\rho
\end{equation}
this is not the case for angular momentum (\ref{defS}). The redshift of $P_\nu
$
under the transformation to accelerated frame $C_\mu $ thus depends on spin
\cite{EPL97} whereas the redshift of mass has a
classical form, as a consequence of transversality (\ref{trans}) of spin
\begin{eqnarray}
\left( C_\mu ,P_\nu \right) &=& 2\left( \eta _{\mu \nu }P^\rho \cdot X_\rho
-P_\mu \cdot X_\nu +P_\nu \cdot X_\mu -S_{\mu \nu }\right)  \nonumber \\
\left( C_\mu ,M^2\right) &=& 4M^2 \cdot X_\mu  \label{CP}
\end{eqnarray}
This mass shift is proportional to mass itself and to the position measured
along the direction of acceleration, which is the form expected for the
potential energy of a mass in a constant gravitational field \cite
{Einstein07}. Einstein redshift law is thus recovered for the quantum shift
of mass written in terms of quantum space-time observables. This property
may in fact be used to define space-time observables \cite{EPL97}. We can
also remind that the shifts of observables $X_\mu $ under transformations to
accelerated frames do no take a classical form, and this is already true for
spinless systems \cite{PLA96}.

The commutators of different position components provide a further
illustration of the changes that have to be performed to shift from a
classical to a quantum description of localisation in space-time. It indeed
results from conformal algebra (\ref{ConfAlg}) that different position
components do not commute except for spinless systems
\begin{equation}
\left( X_\mu ,X_\nu \right) =\frac{S_{\mu \nu }}{M^2}  \label{XX}
\end{equation}
Positions and spin do not commute either
\begin{eqnarray}
\left( S_\mu ,X_\nu \right) =\frac{P_\mu S_\nu }{M^2}
\end{eqnarray}
The commutator (\ref{XX}) clearly indicates that the concepts originating
from classical relativity have to be modified in a quantum
framework. Space-time observables differ from mere coordinate parameters
and frame transformations from changes of coordinate
maps, since the former are directly related to relativistic symmetries and
the latter to conventional parametrizations \cite{Norton93}.

The existence of a covariant definition (\ref{defX}) for space-time
observables is alluded to in the book of
Barut and Raczka \cite{BarutRaczka}.
The physical interest of this definition is questionned because
covariant commutation relations do not seem to be compatible with the
usual laws of motion. The time component $X_0$ cannot be
identified with the time parameter used in standard quantum formalism
and it appears difficult in these conditions to design
a canonical Hamiltonian formalism.
In the context of the present paper, we may notice a further difficulty
raised by the non-commutativity of position components,
namely that it might be uneasy to link the shifts of observables
with the differential laws typical of classical relativity.

To address these difficulties, we first
emphasize once more that the time operator $X_0$ is a localisation
observable, that is precisely the date associated with an event. This
observable clearly differs from any kind of evolution parameter
in particular because the
date associated with an event is a conserved quantity, that is a quantity
preserved by evolution \cite{PLA96}.
More profoundly, we show in the next sections how to bypass
the objections discussed in the previous paragraph.
We first derive laws which generalise the
differential laws of classical relativity.
We then design a canonical representation of localisation
observables which is fully equivalent to the covariant representation.

\section{Quantum conformal factor}

In classical relativity, each map deformation is associated with
a change of the metric tensor.
For conformal transformations, this change reduces to a point-dependent
rescaling which is deduced from the map deformations (\ref{map})
\begin{eqnarray}
\frac{\partial \delta _\nu }{\partial x^\mu }+\frac{\partial \delta _\mu }{%
\partial x^\nu } &=&-2\eta _{\mu \nu }\lambda (x)  \nonumber \\
\lambda _{P_\nu }(x)=0 &\qquad &\lambda _{J_{\nu \rho }}(x)=0  \nonumber \\
\lambda _D(x)=-1 &\qquad &\lambda _{C_\nu }(x)=-2x_\nu  \label{cfac}
\end{eqnarray}
This is because of this conformal character that these
transformations preserve the velocity of light as well
as the propagation of massless fields \cite{BC09}. Notice that the classical
expressions (\ref{cfac}) are written in terms of classical coordinate
parameters and are therefore not properly defined from an operational point
of view. This corresponds to the well known fact that metric factors are not
relativistic observables \cite{Norton93}.

We may however define a conformal factor pertaining to the world
of quantum observables. We indeed see by direct inspection that
the shift of the mass observable under the action of any
conformal generator $\Delta $ may be written
\begin{equation}
(\Delta ,M)= - M \cdot \lambda (X)  \label{Cfac}
\end{equation}
where $\lambda (X)$ is the classical function $\lambda$ evaluated
in terms of a quantum argument given by space-time positions $X$.
Mass is invariant under Poincar\'{e} transformations
while it undergoes the expected space-independent shift
under dilatation. The shift of mass under
transformations to accelerated frames is proportional to the
position measured along the direction of acceleration,
in consistency with Einstein's redshift law.

We show now that this quantum conformal
factor may also be defined from variations under translations
of the shifts of space-time observables,
in conformity with the differential definition (\ref{cfac})
of classical relativity.
To this purpose, we first write that canonical commutators are
classical numbers which commute with any generator $\Delta $
\begin{equation}
\left( \Delta ,\left( P_\mu ,X_\nu \right) \right) =0
\end{equation}
In other words, these commutators are invariant under frame transformations
or, equivalently, the Planck constant is constant like the velocity of light
\cite{Sakharov,Hoyle}, not only under Poincar\'{e} transformations but also
under dilatation and conformal transformations to accelerated frames.
Jacobi identity then entails that space-time shifts $%
\left( \Delta ,X_\nu \right) $ and energy-momentum shifts $\left( \Delta
,P_\mu \right) $ are connected through
\begin{equation}
\left( \left( \Delta ,X_\nu \right) ,P_\mu \right) =\left( \left( \Delta
,P_\mu \right) ,X_\nu \right)  \label{consist}
\end{equation}
Since $\left( \Delta ,X_\nu \right) $ is the shift of position, $\left(
\left( \Delta ,X_\nu \right) ,P_\mu \right) $ is the variation of this shift
under a translation. It is thus a quantum analog of the expression $\frac{%
\partial \delta _\nu }{\partial x^\mu }$ which appears in the classical
equation (\ref{cfac}).

It is possible to strenghten this analogy by
computing the other double commutator $\left( \left( \Delta ,P_\mu \right)
,X_\nu \right) $ which has the roles of positions and momenta interchanged.
For any generator $\Delta $, the momentum shifts $\left( \Delta ,P_\mu
\right) $ are indeed given by conformal algebra (\ref{ConfAlg}) in terms of
translations, rotations and dilatation only. For transformations to
accelerated frames in particular, the shift is read as
\begin{equation}
\left( C_\rho ,P_\mu \right) =2\left( \eta _{\mu \rho }D+J_{\mu \rho }\right)
\end{equation}
We also know that the commutators (\ref{PX}) of translations, rotations and
dilatation generators with the observables $X_\nu $ have a classical form,
so that
\begin{equation}
\left( \left( C_\rho ,P_\mu \right) ,X_\nu \right) =2\left( -\eta _{\mu \rho
}X_\nu -\eta _{\mu \nu }X_\rho +\eta _{\nu \rho }X_\mu \right)
\end{equation}
This is the
classical function which already appears in (\ref{cfac}) but now evaluated
for a quantum argument corresponding to the position observables $X$.
For any generator $\Delta $ more generally, the following results are
obtained
\begin{eqnarray}
\left( \left( \Delta ,P_\mu \right) ,X_\nu \right) &=&-\frac{\partial \delta
_\nu }{\partial x^\mu }\left( X\right)  \nonumber \\
\left( \left( \Delta ,X_\nu \right) ,P_\mu \right) &=&-\frac{\partial \delta
_\nu }{\partial x^\mu }\left( X\right)
\end{eqnarray}
We have used (\ref{consist}) to deduce the second line from the first one.
Quantum analogs of the definition (\ref{cfac}) of the conformal factor are
now obtained by symmetrising the last expressions in the exchange of the two
indices
\begin{eqnarray}
\left( \left( \Delta ,P_\mu \right) ,X_\nu \right) +\left( \left( \Delta
,P_\nu \right) ,X_\mu \right) &=&2\eta _{\mu \nu }\lambda (X)  \nonumber \\
\left( \left( \Delta ,X_\nu \right) ,P_\mu \right) +\left( \left( \Delta
,X_\mu \right) ,P_\nu \right) &=&2\eta _{\mu \nu }\lambda (X)  \label{CFac}
\end{eqnarray}

At this stage, it is worth reminding that the shifts of
space-time and momenta observables under transformations to
uniformly accelerated frames differ from the rules derived from
classical relativity.
Meanwhile the commutators with space-time observables cannot be written
as differential forms because the position components
do not commute in the presence of a non-vanishing spin.
Although these expressions differ from their classical analogs,
they are nevertheless dictated by conformal algebra.
It is in fact a remarkable output of conformal symmetry that the
expressions which determine the metric factor may be brought in
such a simple manner from the classical to the quantum domain.

Moreover, the relativistic transformations of space-time and
energy-momentum scales are now consistently obtained in the quantum domain
from the same commutation relations.
In the particular case $\mu =\nu =0$, the first line of (\ref{CFac}) is
related to variations of redshift of energy-momentum while the second one is
related to variations of shifts of space-time observables. Both relations
are written in terms of the classical function $\lambda $ which
represents the metric factor in classical relativity.
They thus constitute non trivial statements about the
relativistic transformation of the quantum observables associated
with space, time, energy and momentum.
These properties have been derived from conformal symmetry and do not
rely on any further assumption, like the ``clock hypothesis'' of classical
relativity \cite{MTW}. Conformal symmetry is sufficient to
force properly
defined time observables to have their relativistic transformations determined
by the metric factor. To be precise,
this behaviour has been demonstrated here for
transformations to uniformly accelerated frames.
This discussion also means that the conformal
factor may be deduced from measurements of field quantities, although
propagation of electromagnetic field is known to be insensitive to a
conformal variation of the metric tensor.

\section{Canonical representation}

We show in the present section that the covariant formulation of
localisation observables that we have discussed up to now
may be given a canonical representation.

The problem of representing the position of the center of mass of a system
by a quantum operator has been early addressed \cite{Schrodinger,Pryce48}. A
momentum representation of localised quantum states has been used to derive
differential expressions of spin and position consistent with canonical
commutation rules \cite{NewtonWigner}. Operators representing spin and
commuting positions with canonical commutators have also been obtained \cite
{Fleming65}. These definitions, based on purely spatial representations of
quantum observables, with time playing the role of an additional parameter,
were designed to develop a Hamiltonian formalism.
Although these definitions seem to preclude the
possibility of Lorentz invariant commutation relations,
we now derive canonical operators from their covariant
counterparts, showing that both schemes are equivalent.
It will turn out that the canonical formulation not only
involves spatial positions and spin components but also a proper time
operator conjugate to mass.

The canonical spin may be thought as a representation of spin observables in
a center of mass frame \cite{Fleming65}, that is a frame where momentum has
vanishing spatial components. The different spin components can then be
rewritten using the transversality property (\ref{trans})
\begin{equation}
\sigma _{{\rm j}}=S_{{\rm j}}-\frac{P_{{\rm j}}}{P_0+M}S_0\qquad \sigma ^{%
{\rm j}}\sigma _{{\rm j}}=S^\mu S_\mu  \label{sigma}
\end{equation}
or, conversely,
\begin{equation}
S_0=-{\frac{P^{{\rm j}}}M}\sigma _{{\rm j}}\qquad S_{{\rm j}}=\sigma _{{\rm j%
}}+\frac{P_{{\rm j}}}{P_0+M}S_0
\end{equation}
Usual notation $\sigma _{{\rm j}}$ is used for canonical spin variables,
with roman characters denoting spatial indices only. Signs have to be
manipulated with care since $\eta _{{\rm ij}}$ is still used for raising or
lowering indices. The canonical spin components commute with energy-momentum
\begin{equation}
\left( P_\mu ,\sigma _{{\rm j}}\right) =\left( M,\sigma _{{\rm j}}\right) =0
\end{equation}
and their commutation relations may be brought to the usual canonical form
\begin{equation}
\left( \sigma _{{\rm i}},\sigma _{{\rm j}}\right) =-\epsilon _{{\rm ijk}%
}\sigma ^{{\rm k}}\qquad \epsilon _{{\rm ijk}}\equiv \epsilon _{0{\rm ijk}}
\end{equation}

The canonical positions may be understood as the spatial positions of the
center of energy of the system. For spinless systems, the position of center
of energy is well known to be given by the boost generators, that is the
components $J_{0{\rm j}}$ of angular momentum \cite{Einstein06}. For a non
vanishing spin, canonical position components have the generalised form
\begin{equation}
\xi _{{\rm j}}={\frac 1{P_0}}\cdot J_{0{\rm j}}-{\frac{M S_{0{\rm j}}}
{ P_0 \left(P_0+M\right)}}
\end{equation}
Position components have vanishing commutators between themselves as well as
with canonical spin components
\begin{equation}
\left( \xi _{{\rm i}},\xi _{{\rm j}}\right) =0,\qquad \left( \xi _{{\rm i}%
},\sigma _{{\rm j}}\right) =0
\end{equation}
Furthermore, they have canonical commutators with momenta while commuting
with mass
\begin{equation}
\left( P_{{\rm i}},\xi _{{\rm j}}\right) =-\eta _{{\rm ij}},\qquad \left(
M,\xi _{{\rm j}}\right) =0
\end{equation}
The expressions of Poincar\'{e} generators in terms of canonical variables
constitute quantum generalisations of Einstein's relation between Lorentz
boosts and spatial positions \cite{Einstein06}
\begin{eqnarray}
J_{{\rm ij}} &=&P_{{\rm i}}\cdot \xi _{{\rm j}}-P_{{\rm j}}\cdot \xi _{{\rm i%
}}-\epsilon _{{\rm ijk}}\sigma ^{{\rm k}}  \nonumber \\
J_{{\rm 0i}} &=&P_{{\rm 0}}\cdot \xi _{{\rm i}}-\epsilon _{{\rm ijk}}\frac{%
\sigma ^{{\rm k}}P^{{\rm j}}}{P_0+M}  \nonumber \\
P_{{\rm 0}} &=&\sqrt{M^2+P_{{\rm j}}P_{{\rm j}}}
\end{eqnarray}

The previous canonical observables have been obtained from the Poincar\'{e}
generators only. In contrast, the definition (\ref{defX}) of covariant
positions requires the presence of an additional generator, namely the
dilatation $D$. This further generator is necessary to obtain the
localisation time $X_0$. It is also equivalent to the introduction of a
further canonical observable besides the canonical spatial positions
\begin{equation}
D=P^{{\rm j}}\cdot \xi _{{\rm j}}+M\cdot \tau
\end{equation}
The new time operator $\tau $ is seen to commute with all previous
canonical observables, i.e. spatial positions $\xi _{{\rm j}}$, momenta $P_{%
{\rm i}}$ and spin components $\sigma _{{\rm j}}$
\begin{equation}
\left( \tau ,\xi _{{\rm i}}\right) =\left( \tau ,P_{{\rm i}}\right) =\left(
\tau ,\sigma _{{\rm i}}\right) =0
\end{equation}
but to be conjugate to the mass observable
\begin{equation}
\left( M,\tau \right) =-1
\end{equation}
There follows that the covariant positions can be rewritten in
terms of equivalent canonical variables including, besides spatial positions
and momenta, the mass $M$ and the proper time $\tau $
\begin{eqnarray}
X_{{\rm j}} &=&\xi _{{\rm j}}+{\frac{P_{{\rm j}}}M}\cdot \tau +\epsilon _{%
{\rm jik}}\frac{\sigma ^{{\rm k}}P^{{\rm i}}}{M(P_0+M)}  \nonumber \\
X_0 &=&{\frac{P_0}M}\cdot \tau  \label{Xcanon}
\end{eqnarray}
Precisely, mass observable $M$ is the translation operator along the direction
of momentum while $\tau $ is the proper time measured along
the same direction.

We have been able to build a canonical formulation equivalent to the
covariant one. This formulation includes the definition of a canonical
time operator conjugate to the mass observable \cite{Greenberger}.
It allows us to introduce differential notations for representing commutators
with canonical variables
\begin{eqnarray}
\frac{\partial F}{\partial {\xi ^{{\rm j}}}}\equiv -\left( P_{{\rm j}%
},F\right) &\qquad &\frac{\partial F}{\partial {P_{{\rm j}}}}\equiv \left(
\xi ^{{\rm j}},F\right)  \nonumber \\
\frac{\partial F}{\partial \tau }\equiv -\left( M,F\right) &\qquad &\frac{%
\partial F}{\partial M}\equiv \left( \tau ,F\right)  \label{deriv}
\end{eqnarray}
It was not possible to define such differential notations for
covariant observables because of the ambiguities associated with the
non-commutative character of position components.
It is important to emphasize that the proper time is, as the other localisation
observables, a conserved quantity. Hence, equations involving the symbol
$\frac{\partial }{\partial \tau }$ must not be confused with evolution
equations.
They rather represent relativistic shifts in frame transformations
corresponding
to a translation along the direction of momentum.
It is worth keeping this precision in mind when comparing the results obtained
here
with standard Hamiltonian formalism.

As an important application of the canonical representation of observables,
we now introduce velocities as derivatives of space-time observables versus
proper time
\begin{equation}
V_\mu \equiv \frac{\partial X_\mu}{\partial \tau } = - \left( M,X_\mu \right)
\end{equation}
It follows from the canonical commutator (\ref{PX}) that momenta
are related to velocities in a quite simple manner
\begin{equation}
P_\mu = M V_\mu
\end{equation}
The commutation of mass with momenta may then be read
as a quantum version of the law of inertia
\begin{equation}
\frac{\partial P_\mu}{\partial \tau } = M \frac{\partial V_\mu}{\partial \tau }
= M \frac{\partial ^2 X_\mu}{\partial \tau ^2} = 0
\end{equation}
This suggests to reconsider the problem of motion
in a quantum formalism with the help of
the principles of symmetry and, particularly, of
conformal symmetry.

\section{Conclusion}

The localisation of an event is space-time may be characterised by position
quantum observables which transform covariantly under Lorentz
transformations, and which obey Lorentz covariant commutation relations with
momenta. Such position observables have their commutators
determined by the spin components. The Lorentz covariant
observables may be represented equivalently by canonical
operators, at the expense of giving up explicit Lorentz covariance.
Canonical observables satisfy usual canonical commutation relations,
provided a proper time observable conjugate to the mass observable is
included in the canonical representation. The
shifts of covariant observables under
transformations to accelerated frames are closely connected to each other as
well as to the quantum conformal factor.

These results have been obtained as consequences of conformal invariance of
massless field theories. In particular, space-time observables have been
defined for electromagnetic field states consisting in two photons. As
emphasized in the Introduction, these states may be produced by a quantum
event such as the annihilation of an {\em e}$^{+}-${\em e}$^{-}$ pair into a
pair of photons. Energy-momentum variables are conserved in the process and
characterize the {\em e}$^{+}-${\em e}$^{-}$ state as well as the state of
the two emitted photons. It is thus clear that the space-time positions
defined as the point of coincidence of the two electromagnetic pulses
produced by the event, is an image of the space-time position of the
annihilation process. This definition thus enters the general framework of
space-time localisation but it may be applied to massive as well as massless
objects.
As a direct consequence of conservation laws, the conformal variation of
mass which has been established for electromagnetic field states has
to hold also for massive objects. In particular, the mass of elementary
particles
such as {\em e}$^{-}$ or {\em e}$^{+}$ has to vary as the reciprocal of the
space-time scale under conformal transformations to accelerated frames.

We are finally led to question not only the representation of localisation in
space-time but also the representation of movement inherited from classical
physics. To illustrate this point, we consider again the annihilation of an
{\em e}$^{+}-${\em e}$^{-}$ pair and we assume that the two particles are
bound to each other as a positronium atom moving along a given direction.
The annihilation process may occur at different positions in time, each of
them corresponding to a different space-time position in the theoretical
framework of the present paper. The proper time $\tau $ has been defined as
a localisation observable, the variation of which describes a translation
along the direction of motion. Different values of the proper time variable
may therefore be used to distinguish between the different events
which may occur along the same trajectory.
The commutator of any observable with $M$
may be understood as the derivative of this observable with respect
to $\tau$. In particular, a quantum form of the law of inertia
is derived in this manner from conformal symmetry.

This leads to a new conception of motion which is now
apprehended as a collection of events corresponding to different
positions in space-time. Bertrand Russell has lucidly
analysed this `atomistic' conception of space-time
which is one of the most profound conceptual
implications of relativistic theories \cite{Russell25}

\begin{quote}
The world which the theory of relativity presents to our imagination is not
so much a world of `things' in `motion' as a world of {\it events}.
\end{quote}

Such a conception is demanded not only by
relativistic arguments discussed by Russell, but also by the necessity of
defining well behaved quantum observables. In our
opinion, it is quite remarkable that the conformal symmetry, a natural
extension of Lorentz symmetry, allows to lay down the foundations of a
theoretical framework which has the ability of dealing satisfactorily with
relativistic as well as quantum requirements.

\end{multicols}

\end{document}